\documentclass[a4paper,10pt,journal]{IEEEtran}
\usepackage{cite}
\usepackage{siunitx}
\DeclareSIUnit \GHz {GHz}
\DeclareSIUnit \dBm {dBm}
\usepackage{graphicx}
\usepackage{algorithm}
\usepackage{amsfonts}
\usepackage{mathtools}
\usepackage{multirow}
\usepackage{amsmath}
\usepackage[caption=false,font=footnotesize]{subfig}
\usepackage[hyphens]{url}
\usepackage[noend]{algpseudocode}

\newcommand{\FF}[1]{{\mathbb{F}}}

\usepackage{tikz}
\usetikzlibrary{arrows,automata}

\usepackage{enumerate}
\newtheorem{assumption}{Assumption}[section]

\newtheorem{lemma}{Lemma}[section]
\newtheorem{theorem}{Theorem}[section]

\newtheorem{remark}{Remark}[section]

\newcolumntype{P}[1]{>{\centering\arraybackslash}p{#1}}
\newcolumntype{M}[1]{>{\centering\arraybackslash}m{#1}}

\newcolumntype{L}[1]{>{\raggedright\let\newline\\\arraybackslash\hspace{0pt}}m{#1}}
\newcolumntype{C}[1]{>{\centering\let\newline\\\arraybackslash\hspace{0pt}}m{#1}}
\newcolumntype{R}[1]{>{\raggedleft\let\newline\\\arraybackslash\hspace{0pt}}m{#1}}

\DeclareMathOperator*{\argminA}{arg\,min}

\begin{document}
\title{{\LARGE Optimized Split Computing Framework for Edge and Core Devices}}

\author{Andrea Tassi, Oluwatayo Yetunde Kolawole, Joan Pujol Roig, and Daniel Warren
\thanks{
All the authors are with  Communication Solutions at Samsung R\&D Institute UK (SRUK), UK (e-mail: {\tt \{a.tassi, o.kolawole, j.pujolroig, dan.warren\}@samsung.com}). This work is a contribution by Project REASON, a UK Government funded project under the Future Open Networks Research Challenge sponsored by the Department of Science Innovation and Technology.
}}

\maketitle
\begin{abstract}
With mobile networks expected to support services with stringent requirements that ensure high-quality user experience, the ability to apply Feed-Forward Neural Network (FFNN) models to User Equipment (UE) use cases has become critical. Given that UEs have limited resources, running FFNNs directly on UEs is an intrinsically challenging problem. This letter proposes an optimization framework for split computing applications where an FFNN model is partitioned into multiple sections, and executed by UEs, edge- and core-located nodes to reduce the required UE computational footprint while containing the inference time. An efficient heuristic strategy for solving the optimization problem is also provided. The proposed framework is shown to be robust in heterogeneous settings, eliminating the need for retraining and reducing the UE's memory (CPU) footprint by over $33.6\%$ ($60\%$).
\end{abstract}


\section{Introduction}\label{sec:intro}
 The technological advancements in mobile communications towards 6G have driven an upsurge in the incorporation of artificial intelligence (AI) and machine learning (ML) applications in networks. Thus, deep neural networks are being employed to optimize tasks and execute complex inferences to support the demands for unprecedented quality of user experiences as well as enabling zero-latency connectivity. Moreover, the expectation is for the full execution of complex tasks across the network devices, even on edge devices such as smartphones, IoT devices, etc.
 
 However, the surge in predictive accuracy for real-time operations of modern FFNN models is accompanied by a spike in computational demands and energy consumption~\cite{Tan2020}. Given smartphones' inherent power and computational limitations, this makes deploying FFNN models in mobile applications particularly daunting. Hence, the need for a mobile device to partially offload its FFNN model inference tasks onto other edge- or cloud-located processing nodes~\cite{HexaX6G}. This letter focuses on mitigating the computational and memory footprint associated with running inference tasks on pre-trained Feed-Forward Neural Networks (FFNNs) from low-end User Equipments (UEs) onto other more resourceful network entities. 

A prevalent approach in the industry is offloading the entire computation and processing tasks to edge and/or cloud-based processing nodes~\cite{Wang2020,Tassi,10944600}. Although alleviating the computation burden, these edge-/cloud-based computing methods do not intrinsically consider the quality of wireless links. These connections are susceptible to erratic disruptions due to unpredictable interference and noise patterns of wireless channels. To this end, some works advocate creating bespoke, lightweight FFNN models tailored for resource-limited mobile devices~\cite{Liang2021}. Yet, this is not always feasible and often compromises the models' accuracy, resulting in a diminished quality of predictions and inference.  

Hybrid strategies like Split Computing (SC) have emerged~\cite{Bakhtiarnia2023, Matsubara2022}. Originally, SC was a technique designed to partition neural network models into two sections: the head, which operates at the UE, and the tail, processed at an edge server. The head section handles initial data processing, while the tail section manages deeper analyses at the resource-rich edge server. A unique feature of SC is its dynamic splitting point, allowing the model to adjust in real-time to both fluctuating wireless channel conditions and the specific constraints of the UE and/or cloud server. This adaptability ensures efficient data transfer, optimized performance, and reduced energy consumption, making SC a pivotal solution for deploying neural network models in resource-constrained environments. 

This letter expands and builds upon the current state-of-the-art (see~\cite{Liang2021,Bakhtiarnia2023,Matsubara2022} and the references therein) with the following novel contributions:
    \newline
    \textbf{End-to-End Optimization:} We propose a framework for determining an arbitrary number of splitting points while taking into account the link quality between the various devices running inference tasks and their computational capabilities. In the considered system model, UEs always initiate FFNN inference tasks. Given such tasks' substantial computational and memory demands, the inference workload is distributed among the initiating UE, its associated base station, and, where applicable, a core network processing node.
    \newline
    \textbf{Costs Reduction:} The proposed optimization model accounts for computational and memory resources available across all the network actors that may play a part in completing an FFNN inference task and minimizes the communication delay associated with the network links interconnecting the actors.
    \newline
    \textbf{Dynamic Adaptation:} Our proposed framework provides an efficient heuristic strategy for calculating feasible solutions to the problem, eliminating the necessity of retraining a model when significant system changes occur.

The rest of the paper is structured as follows. Section~\ref{sec:systemModel} presents the considered system model. The proposed SC optimization problem and the procedure to calculate its heuristic solution are described in Sections~\ref{sec:opt} and~\ref{sec:procedure}, respectively. A performance evaluation of the proposed approach is given in Section~\ref{sec:evaluation}. Finally, in Section~\ref{sec:conclusion}, we draw our conclusions.

\section{System Model}\label{sec:systemModel}
We refer to a system model consisting of a set of devices $\mathcal{D} = \{d_1, d_2,\ldots, d_{|\mathcal{D}|}\}$ capable of running (partially or in full) an inference in a given FFNN model. The computational capacities of each device are listed in set $\mathcal{R} = \{(r^{(\textrm{CPU})}_i, r^{(\mathrm{Mem})}_i), \forall i \in [|\mathcal{D}|]$ \footnote{For positive integer $K$, $[K]$ denotes the set $\{1,2,\dots, K\}$}, where tuple $(r^{(\textrm{CPU})}_i, r^{(\mathrm{Mem})}_i)$ delineates the peak CPU capacity and the utmost memory allocation for device $d_i$, respectively. 
\par 
CPU capacities are articulated as an aggregate of individual capacities, with $r_i^{(\textrm{CPU})} \in [0,1]$, representing the allocation on each CPU core. This is then normalized concerning the highest CPU capacity across all devices. Similarly, memory capacities are normalized based on the maximum memory allocation of any device, thus $r_i^{(\mathrm{Mem})} \in [0,1]$.
\par 
The connectivity blueprint of these devices, detailing both wired and wireless interconnections, is represented by a matrix $\mathbf{W}$ of dimensions $|\mathcal{D}| \times |\mathcal{D}|$. In this matrix, the element $w_{i,j}$ is assigned a value of $0$ if there is no network link between device  $d_i$ and device $d_j$. If a link exists, the maximum data rate is assigned to this link (measured in \SI{}{bit/s}). If $d_i$ and $d_j$ communicate via a wireless link, $w_{i,j}$ is a direct function of the Modulation and Coding Scheme (MCS) associated with the device $d_i$'s uplink communications, which is a high-level quantized indication of the propagation condition affecting the link. It is beyond the scope of the paper to focus on specific network technologies. Yet, modern 4G/5G cellular networks operate on the assumption of continuously gathering network quality information among all the wired-/wirelessly-connected actors, making it straightforward to populate  $\mathbf{W}$~\cite{5GLENA}. We now formulate the following assumption:
\begin{assumption}\label{ass.1}
The network interconnecting each device may be partitioned into a number of sub-networks, where a device may communicate with another pertaining to the same or a different sub-network. For the sake of compactness in our notation, we say that a \emph{single logical network hop} exists between devices $d_i$ and $d_j$ (for $i \neq j$ and $i,j \in [|\mathcal{D}|]$) if $d_i$ and $d_j$ belong to the same or different sub-networks, and $d_i$ communicates with $d_j$ via a number of switches and/or routers (or none). Furthermore, a single logical network hop exists between $d_i$ and $d_j$ if and only if $j - i = 1$. For any pair of devices $d_t$ and $d_u$ (where $u-t > 1$, $t < |\mathcal{D}|$ and $u < |\mathcal{D}|$), $d_t$ can communicate with device $d_u$ only via the sequence of intermediate devices $[d_{t+1}, d_{t+2}, d_{t+3}, \ldots, d_{u-1}]$.
\end{assumption}

The aforementioned assumption introduces a pipeline-like way of communicating among the devices involved in the distributed execution of an FFNN model inference task. This directly follows from the fact that it is impossible to parallelize the execution of non-overlapping sets of FFNN model layers over multiple devices. Without loss of generality, we make the following assumption:
\begin{assumption}\label{ass.3}
Only device $d_1$ is the device capable of initializing an inference task.
\end{assumption}

{The proposed generalized device hierarchy mimics the structure of modern mobile cellular networks where UEs are (largely) only allowed to communicate with the base station they are connected to. The proposed modeling framework will need to be extended to make it applicable to system models relying upon an arbitrary device connectivity graph (e.g., mesh-like connectivity graphs).}

We model a FFNN model as a tuple $(\mathcal{L},\mathcal{C},\mathbf{B})$. Set $\mathcal{L} = \{l_1, l_2\ldots, l_{|\mathcal{L}|}\}$ consists of the FFNN model's layers, where $l_i$ ( $\forall i \in [|\mathcal{L}|]$) denotes the $i$-th layer in the model.
\par
Execution costs of each layer are listed in set $\mathcal{C} = \{(c^{(\textrm{CPU})}_i, c^{(\mathrm{Mem})}_i), \forall i \in [|\mathcal{L}|]\}$, where tuple $(c^{(\textrm{CPU})}_i, c^{(\mathrm{Mem})}_i)$ specifies the CPU cost and the memory cost for layer $\ell_i$, respectively. Both CPU and memory costs are normalized using the previously mentioned factor.
Data generated by each layer during a single FFNN model inference is detailed in the
 matrix  $|\mathcal{L}| \times |\mathcal{L}|$ matrix $\mathbf{B}$. Here, the element  $b_{i,j}$ represents the overall amount of data  (expressed in \SI{}{\bit}) traversing connections from layer $l_i$ and terminating to layer $l_j$. If no connections exist from $l_i$ to $l_j$, then $b_{i,j}$ is set equal to $0$.
\par
To align with contemporary FFNN models, we adopt the subsequent assumption:
\begin{assumption}\label{ass.2}
For any layer pair $l_i$ and $l_j$ (where $i,j \in [|\mathcal{L}|]$), if $i < j$, then $b_{i,j} \geq 0$. On the other hand, if $i \geq j$, then $b_{i,j} = 0$.
\end{assumption}

\section{Proposed Optimization Framework}\label{sec:opt}
We introduce integer optimization variables, denoted by $\mathbf{x} = [x_1, \ldots, x_{\kappa}]$, where $\kappa$ ranges from $1$ to $\hat{\kappa}$. Each variable represents a splitting point in an FFNN model. The maximum number of splitting points is $1 \leq \hat{\kappa} \leq |\mathcal{L}|$. The possible values for each optimization variable are defined as:
\begin{equation}\label{eq.k}
  x_t =
  \begin{cases}
	 \{1, \ldots, (|\mathcal{L}|-1)\} & \text{for } t = 1, \ldots, (\kappa-1)\\
	 \{|\mathcal{L}|\} & \text{for } t = \kappa.
  \end{cases}
\end{equation}
The optimization variables partition the set $\mathcal{L}$ into non-overlapping subsets $\mathcal{L}^{(1)}, \ldots, \mathcal{L}^{(\hat{\kappa})}$ as:
\begin{equation}\label{eq.L}
  \hspace{-2mm}\mathcal{L}^{(t)} =
  \begin{cases}
	 \{1, 2,\ldots, x_1\} & \hspace{-1mm}\text{for } t = 1\\
	 \{x_{t-1} + 1, x_{t-1} + 2,\ldots, x_t\} & \hspace{-1mm}\text{for } t = 2, \ldots, \kappa.
  \end{cases}
\end{equation}
It is evident that $\bigcup_{t = 1}^{\hat{\kappa}} \mathcal{L}^{(t)} \doteq \mathcal{L}$. From relation~\eqref{eq.L}, we deduce that $x_1 < x_2 < \ldots < x_{\hat{\kappa}}$.

Considering Assumption~\ref{ass.3}, for notation simplicity, we assume that FFNN model layers in set $\mathcal{L}^{(t)}$ map onto device $d_t$ for $t = 1, \ldots, \hat{\kappa}$. Thus, $\hat{\kappa}$ equals $\min\{|\mathcal{D}|, |\mathcal{L}|\}$.

The constraints below ensure that the CPU cost of the FFNN model layers in $\mathcal{L}^{(t)}$ does not surpass the CPU capacity of device $d_t$:
\begin{align}
  \max_{i \in \mathcal{L}^{(t)}} \left\{c^{(\textrm{CPU})}_i\right\} \leq r^{(\textrm{CPU})}_t, \quad \forall t \in [\hat{\kappa}]. \label{OM.c1}
\end{align}
Each FFNN model layer's inference task can be divided into sub-tasks for each layer, which are run sequentially

Similarly, we ensure that the total memory footprint of the FFNN model layers in $\mathcal{L}^{(t)}$ is within the memory capacity of device $d_t$:
\begin{align}
  \sum_{i \in \mathcal{L}^{(t)}} c^{(\mathrm{Mem})}_i \leq r^{(\mathrm{Mem})}_t, \quad \forall t \in  [\hat{\kappa}]. \label{OM.c2}
\end{align}
All FFNN model layers on a device are pre-loaded in its memory, irrespective of the inference sub-task status. Thus, device $t$ must have sufficient memory to store layers in $\mathcal{L}^{(t)}$.

Considering a system with three devices ($|\mathcal{D}| = 3$) and some FFNN model layers mapped to device $d_1$ connected to layers on device $d_3$, due to Assumption~\ref{ass.1}, the network link between device $d_1$ and device $d_2$ must accommodate data streams from the first to the second and third devices. Following this logic, we define the objective function:
\begin{align}
\Psi(\mathbf{x};\kappa) = \sum_{t = 1}^{\kappa - 1}\overbrace{\frac{1}{w_{t, t+1}} \quad \,\sum_{\mathclap{\substack{i \in \bigcup_{h = 0}^{t} \mathcal{L}^{(h)}, \\ j \in \bigcup_{u = t+1}^{\kappa} \mathcal{L}^{(u)} }}} b_{i,j}}^{\psi^{(x_t)}}, \label{OM.of}
\end{align}
where term \(\psi^{(x_t)}\) refers to the total data volume from device \(d_t\) to devices \(d_{t+1}, \ldots, d_\kappa\), normalized by the network link bandwidth between devices \(d_t\) and \(d_{t+1}\). Specifically, \(\psi^{(x_t)}\) denotes the time required to transfer the entire data stream generated in a single FFNN inference from device \(d_t\) to device  \(d_{t+1}\). For convenience, we define \(\psi^{(x_\kappa)} = 0\). Notably, \eqref{OM.of} only accounts for the data stream originating from layer partition \(d_t\) and directed to layers in another partition with an index $i\geq t$. This aligns with Assumption~\ref{ass.2}, which restricts an FFNN model \(\ell\) from connecting to layers with indexes less than or equal to \(\ell\) (for \(\ell = 1, \ldots, |\mathcal{L}|\)). We assume the time to transfer the output of an FFNN inference back to device \(d_1\) is negligible, and thus, it is not considered in our model.

For a specified number of splitting points \(\kappa\), the Split Computing Optimization (SCO-\(\kappa\)) problem is defined as:
\begin{align}
  \text{SCO-}\kappa: &  \quad  \min_{\mathbf{x}} \,\,  \Psi(\mathbf{x};\kappa), \label{OM.of_}\\
\text{subject to} &  \quad\eqref{OM.c1} \text{ and } \eqref{OM.c2}. \label{OM.c0_}
\end{align}
Denote \(\mathbf{\hat{x}}^{(\kappa)}\) as the optimal solution of the SCO-\(\kappa\) problem. 
We prioritize split computing solutions with the fewest splitting points. Thus, the globally optimum split computing solution \(\mathbf{x}^*\) for the SCO problem is:
\begin{align}
\text{SCO} \quad \argminA_{\kappa = 1, \ldots, \hat{\kappa}}\left\{\Psi(\mathbf{\hat{x}}^{(\kappa)};\kappa)\right\}. \label{OM.Global}
\end{align}

\section{Proposed Split Optimization Procedure}\label{sec:procedure}
\begin{algorithm}[t]
\floatname{algorithm}{Procedure}
 \begin{algorithmic}[1]
\caption{Split Optimization Procedure.}\label{alg:SOP}
\begin{scriptsize}
\Procedure{$\textsc{SO}(\Hat{\kappa})$}{}
    \For {$\kappa = 1, \ldots, \Hat{\kappa}$}
   	 \State $x_1 \gets 0$
   	 \State Initialize: \(\epsilon \gets 1\), \(\delta \gets 0\), \(\ell \gets 1\), \(\nu \gets \text{\textbf{true}}\), \(\psi \gets [\infty, \ldots, \infty]\)
   	 \While {$\ell \leq |\mathcal{L}|$}
   		 \If {Conditions \(c^{(\text{CPU})}_\ell \leq r^{(\text{CPU})}_\epsilon\) and \(\delta + c^{(\text{Mem})}_\ell \leq r^{(\text{Mem})}_\epsilon\) are met}
       		 \State $x_\epsilon \gets x_\epsilon + 1$
                  \State $\psi[\ell] \gets \psi^{(x_\epsilon)}$
                  \State $\delta = \delta + c^{(\text{Mem})}_\ell$
                  \State $\ell \gets \ell + 1$
   		 \Else
                    \If {$\epsilon + 1 > \kappa$} 
                        \State $\nu \gets \text{\textbf{false}}$
                        \State \textbf{break}
                    \Else
                        \State $x_\epsilon \gets \argminA\{\psi\}$
                        \State $\epsilon \gets \epsilon + 1$
                        \State $x_\epsilon \gets x_{\epsilon-1}$, $\ell \gets x_{\epsilon-1}$, $\psi \gets [\infty, \ldots, \infty]$, $\delta \gets 0$
                    \EndIf
   		 \EndIf
   	 \EndWhile
          \State \textbf{If} Solution is valid (\(\nu == \text{\textbf{true}}\)) \textbf{then} \textbf{return} \(\textbf{x}\)
    \EndFor
    \State \textbf{return} No valid solution found.
\EndProcedure
\end{scriptsize}
\end{algorithmic}
\end{algorithm}

\AddToHookNext{shipout/background}{
  \begin{tikzpicture}[remember picture,overlay]
  \draw[fill=gray, fill opacity=0.2, rounded corners, draw=none] (10.2,-7.9) rectangle (2.5,-2.95);
  \node[draw=none,align=left, rotate=90] at (2.3,-5.4) {\emph{{\small \color{black}SCO-$\kappa$ Solving Task}}};
  \end{tikzpicture}
}

We introduce the Split Optimization (SO) procedure to heuristically compute a feasible solution for the SCO at~\eqref{OM.Global}. The heuristic's core steps are outlined in Procedure~\ref{alg:SOP} where each iteration of the for-loop (lines 2-19) attempts to compute a feasible solution for an instance of the SCO-\(\kappa\) problem for \(\kappa \in [\Hat{\kappa}]\) (\emph{SCO-\(\kappa\) solving task}). Upon finding a solution for one of the SCO-\(\kappa\) problems, Procedure~\ref{alg:SOP} deems that solution valid for the SCO problem and returns (line 19).

A SCO-\(\kappa\) solving task attempts to solve one instance of the SCO-$\kappa$ problem. In particular, the while-loop (lines 5-18) begins by setting the first splitting point in an FFNN model equal to the first layer ($x_1 = 1$). The value of $x_1$ is progressively incremented for as long as constraints~\eqref{OM.c1} and~\eqref{OM.c2} are met (lines 7-10). When any or both the aforementioned constraints are about to be violated, $x_1$ is set equal to the FFNN layer index minimizing the term $\psi^{(x_t)}$, for $t = 1, \ldots, \kappa$ (line 16). Then, the while-loop repeats by considering the next splitting point ($x_2$), which can only take values greater than $x_1$ (lines 18 and 7). The while-loop terminates as soon as all the FFNN model layers have been considered (namely, $\ell > |\mathcal{L}|$) or the index $\epsilon$ of the currently considered splitting point exceeds $\kappa$. It is immediate to observe that the while-loop iterates for no more than $|\mathcal{L}| + \kappa - 1$ times.

The following lemma demonstrates that Procedure~\ref{alg:SOP} yields feasible solutions for the SCO-\(\kappa\) problem.
\begin{lemma}\label{lemma.1}
Given \(\kappa \in [\Hat{\kappa}]\), if the while-loop in lines 5-18 of Procedure~\ref{alg:SOP} concludes due to the violation of the loop condition at line~5, then the set \(\{x_1, \ldots, x_\kappa\}\) is a feasible solution for the corresponding SCO-\(\kappa\) problem.
\end{lemma}
\begin{IEEEproof}
The proof is derived from the fact that the while-loop in lines 5-18 of Procedure~\ref{alg:SOP} minimizes the terms \(\psi^{(x_t)}\) for \(t = 1, \ldots, \kappa - 1\). During this minimization, line~6 ensures constraints~\eqref{OM.c1} and~\eqref{OM.c2} are met.
\end{IEEEproof}
\par
From Lemma~\ref{lemma.1}, Theorem~\ref{th.1} directly follows.
\begin{theorem}\label{th.1}
Let \(\overline{\mathbf{x}} = \{x_1, \ldots, x_{\overline{\kappa}}\}\) be a solution returned by Procedure~\ref{alg:SOP}, where \(\overline{\kappa} \leq \Hat{\kappa}\). Then, \(\overline{\mathbf{x}}\) is a feasible solution for the SCO problem.
\end{theorem}
We also formulate the following remarks.
{\begin{remark}\label{rem.1}
The SO procedure establishes split computing solutions associated with the smallest number of splitting points $\kappa$ that minimizes~\eqref{OM.of}. As such, should a device subset $\left\{d_i\right\}_{i = 1}^{\Tilde{D}} \subset \mathcal{D}$ (for $\Tilde{D}$ being a non-negative integer smaller than $|\mathcal{D}|$) be associated with enough CPU and memory capacity to accommodate an FFNN model, the SO procedure will return a split computing solution with $\kappa < |\mathcal{D}| - 1$, i.e., a split computing solution mapping the FFNN model's layers on fewer devices than $|D|$.
\end{remark}}

{\begin{remark}\label{rem.2}
The entirety of vector-based operations of the SO procedure takes place within the body of the while-loop of the \emph{SCO-\(\kappa\) solving task}. Hence, a qualitative indication of the SO procedure's computational complexity is a direct function of the number of times the aforementioned while-loop iterates (worst case) and the complexity of the operations the while-loop consists of. As for the latter, all the operations amount to element-by-element summations, products, and inversions involving short vectors (no longer than $|\mathcal{D}| - 1$), with the exception of a single $\arg\min$ (line 16) that is still calculated over a short vector (with the same length indicated above). Modern CPU architectures and scientific programming tools make the execution of vector-wise basic arithmetic operations very time-efficient. Overall, we already observed that, for a given $\kappa$, the while-loop of the \emph{SCO-\(\kappa\) solving task} iterates for no more than $|\mathcal{L}| + \kappa - 1$ times. Hence, during the execution of the SO procedure, the aforementioned while-loop iterates for no more than $\zeta$ times, defined as follows:
\begin{equation}
\zeta = \sum_{\kappa = 1}^{\Hat{\kappa}}\left(|\mathcal{L}| + \kappa - 1\right) = \frac{\Hat{\kappa}^2}{2} + \frac{2|\mathcal{L}| - 1}{2}\Hat{\kappa}.
\end{equation}
In the considered simulation scenarios (see Sec.~\ref{sec:evaluation}), the largest values of $|\mathcal{L}|$ and $\Hat{\kappa}$ are set equal $517$ and $5$, respectively. As such, during the execution of the SO procedure, the while-loop of the \emph{SCO-\(\kappa\) solving task} iterates for no more than $2595$ times.
\end{remark}}

\section{Analytical Results}\label{sec:evaluation}
In this section, we refer to an equivalent linear formulation of the SCO-$\kappa$ problem that can be obtained by adopting, after some manipulations, the transformations presented in~\cite[Eqs. (4)-(7), (11)-(14), (78)-(81)]{math10020283}. In doing so, the transformed SCO-$\kappa$ problem is an integer linear problem equivalent to the formulation as per~\eqref{OM.of_}-\eqref{OM.c0_}.

\begin{figure}[t]
\centering
\hspace*{-1.5mm}\subfloat[Average cost difference vs. $|\mathcal{L}|$]{\label{fig.1.1}
    \includegraphics[width=0.5\columnwidth]{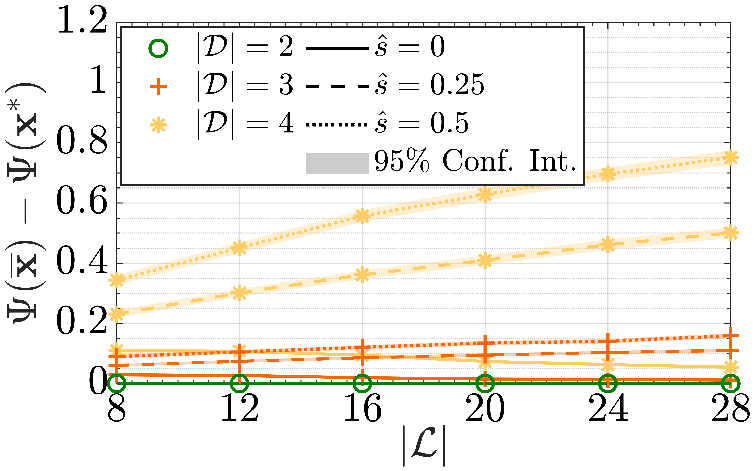}
}
\hspace*{-1.8mm}\subfloat[Average cost difference vs. $|\mathcal{D}|$]{\label{fig.1.2}
    \includegraphics[width=0.5\columnwidth]{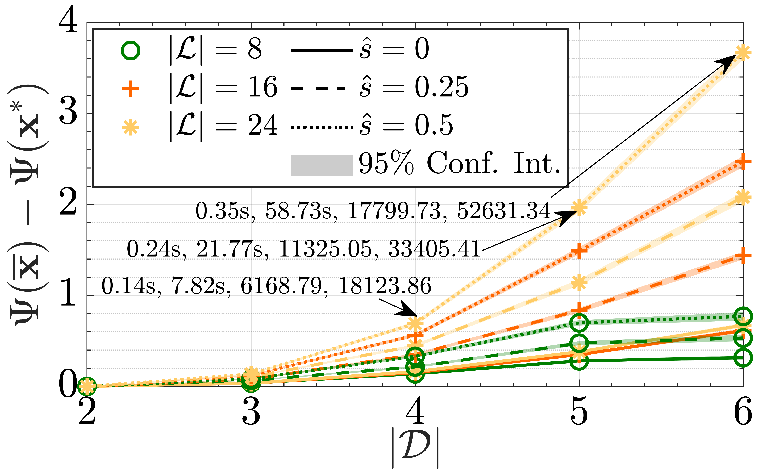}
}
\caption{Normalized average cost difference as a function of the total number of FFNN model layers and devices; $95\%$ confidence intervals are also included. Each annotated point lists the average SO procedure and SCIP completion times (measured on a workstation equipped with a CPU AMD 5995WX with cores operated at $\SI{1.8}{\giga\hertz}$), the average number of optimization variables and constraints of the linearized version of the SCO problem.}
\label{fig.1}
\end{figure}

We optimally solved the SCO problem by calculating the optimum solution of an equivalent linear formulation of the SCO-$\kappa$ problem, for $\kappa \in [\Hat{\kappa}]$ by means of the Solving Constraint Integer Programs (SCIP) solver~\cite{BestuzhevaEtal2021ZR}. This makes it straightforward to establish $\mathbf{x}^*$ by directly solving~\eqref{OM.Global}. We remark that $\Psi(\mathbf{x}^*)$ and $\Psi(\mathbf{\overline{x}})$ are the values of the objective function~\eqref{OM.of} when the SCO problem is optimally solved and when the solution is obtained via the proposed SO procedure, respectively. Since $\Psi(\overline{\mathbf{x}}) \geq \Psi(\mathbf{x}^*)$, Fig.~\ref{fig.1} shows how much a heuristic solution $\mathbf{\overline{x}}$ deviates from the corresponding optimal solution $\mathbf{x}^*$ by showing the \emph{cost difference} $\Psi(\mathbf{\overline{{x}}}) - \Psi(\mathbf{x}^*)$ as a function of the number of an FFNN model's layers $|\mathcal{L}|$ and available devices $|\mathcal{D}|$.

To effectively investigate the performance of the proposed SO procedure, in Fig.~\ref{fig.1}, we considered FFNN models with a number of layers $|\mathcal{L}| \in [8,28]$. Given an FFNN model, each layer $\ell_i$ is associated with a normalized CPU footprint $c_i^{\textsc{(CPU)}}$ equal to $1$, for $i \in [|\mathcal{L}|]$ -- we regard $\overline{C}$ as the CPU normalization factor that is set equal to the largest CPU capacity among all the devices in $\mathcal{D}$. The normalized memory footprint of the aforementioned FFNN model's layer $c_i^{\textsc{(Mem)}}$ is chosen uniformly distributed at random in $(0.01, 1.0]$ -- we set the memory normalization factor $\overline{M}$ equal to the largest memory capacity among all the devices in $\mathcal{D}$. Matrix $\mathbf{B}$ is a direct function of the FFNN model's layer memory footprint. In particular, 
the number of bits $b_{i,j}$ to be transferred from layer $\ell_i$ to layer $\ell_{j}$, for $1 \leq i < j \leq |\mathcal{L}|$, is set equal to $c_i^{\textsc{(Mem)}}$. For $j \geq i + 1$ and $j < |\mathcal{L}|$, we considered a probability $\hat{s} \in \{0, 0.25, 0.5\}$ of $b_{i,j}$ being non-zero. For a given number of devices and FFNN model's layers, we considered $\overline{I} = 10^4$ Monte Carlo iterations.
These simulation settings consist of a variable number of devices $|\mathcal{D}| \in [2,6]$ each offering a normalized CPU capacity $r_i^{\textsc{(CPU)}}$, and a normalized memory capacity $r_i^{\textsc{(Mem)}}$ defined as follows:
\begin{equation}
    r_i^{\textsc{(CPU)}} = \sum_{i \in [\mathcal{L}]} \frac{c^{\textsc{(CPU)}}_i}{ |\mathcal{D}| - i + 1},
\end{equation}
\begin{equation}
    r_i^{\textsc{(Mem)}} = \sum_{i \in [\mathcal{L}]} \frac{c^{\textsc{(Mem)}}_i}{|\mathcal{D}| - i + 1}.
\end{equation}
For instance, if $|\mathcal{D}| = 2$ then $d_2$ is associated with $r_2^{\textsc{(Mem)}}$ equal to the sum of the FFNN model's layer memory footprints, and $d_1$ is associated with $r_1^{\textsc{(Mem)}}$ that is only half of $r_2^{\textsc{(Mem)}}$. The normalized bandwidth $w_{t,u}$ of the wireless/wired link interconnecting devices $d_t$ and $d_u$, for $1 \leq t < u \leq |\mathcal{D}|$, is set equal to $1 / (|\mathcal{D}| - 1)$ -- the bandwidth normalization factor $\overline{B}$ is set equal to largest bandwidth value among all the links. These modeling assumptions capture edge devices' reduced memory capacities and communication bandwidths.

Fig.~\ref{fig.1.1} shows the average cost difference (averaged across all the randomly generated instances of $(\mathcal{L}, \mathcal{C}, \mathbf{B})$) as a function of the number of FFNN models' layers. For a number of devices $|\mathcal{D}| = 2$, we observe that the average cost difference is equal to zero, regardless of the considered number of FFNN model's layers and the value of $\hat{s}$ -- thus, signifying the proposed SO procedure returns a solution associated with the same cost of the solution returned by employing the SCIP solver. As the number of devices increases to $|\mathcal{D}| = 3$, the average cost difference tends to remain stable for a given value of $\hat{s}$. As the value of $\hat{s}$ increases, the average cost difference also tends to increase. For instance, this can be observed for $\hat{s} = 0.5$, when $|\mathcal{L}|$ increases from $8$ to $28$, the average cost difference increases from $0.089$ to $0.158$. At the same time, for $|\mathcal{L}| = 28$, as the value of $\hat{s}$ increases from $0$ to $0.5$, the average cost difference increases from $0.011$ to $0.158$. Overall, if $|\mathcal{D}| = 3$, the value of $\hat{s}$ impacts the average cost difference more than the total number of FFNN model layers. This is not surprising as, for a relatively small number of devices in the system model, the structure of matrix $\mathbf{B}$ determines the complexity of the SCO problem. However, as soon as the number of devices increases to $4$, both the number of FFNN model layers and the value of $\hat{s}$ have an impact on the overall complexity of the SCO problem, which results in average cost differences that increases as the values of $|\mathcal{L}|$ and/or $\hat{s}$ increases. If we consider the case where $\hat{s}$ is set to its largest value ($0.5$), the average cost difference moves from $0.343$ (for $|\mathcal{L}| = 8$) to $0.751$ (for $|\mathcal{L}| = 28$).

\begin{table}
\scriptsize
\renewcommand{\arraystretch}{1.3}
\caption{Main Simulation Parameters 
}
\label{tab.I}
\centering
\begin{tabular}{|c|ccl|C{3cm}|}
\hline
\multirow{3}{*}{\rotatebox[origin=c]{90}{\hspace{-0.55cm}\textbf{FFNN Models}}}                                 & \multicolumn{1}{c|}{Name}                           & \multicolumn{2}{c|}{$|\mathcal{L}|$} & \hspace{-0.5cm}Trainable Parameters $\times \cdot 10^{-7}$                                                                                                                                                                               \\ \cline{2-5} 
                                                                      & \multicolumn{1}{M{1.5cm}|}{R50, R101, R152}                 & \multicolumn{2}{c|}{$177$, $347$, $517$}           & \hspace{-0.5cm}$2.56, 4.46, 6.03$                                                                                                                                                                                                                 \\ \cline{2-5} 
                                                                      & \multicolumn{1}{M{1.5cm}|}{Y1, Y2, Y3, Y4}                    & \multicolumn{2}{c|}{$77$, $100$, $297$, $510$}            & \hspace{-0.5cm}$5.99, 5.08, 6.15, 6.44$                                                                                                                                                                                                                  \\ \hline
\hline
\multicolumn{1}{|l|}{\multirow{7}{*}{\rotatebox[origin=c]{90}{\textbf{Communication Network \hspace{0.3cm}}}}} & \multicolumn{3}{R{3cm}|}{Deployment type}                                                       & \multicolumn{1}{p{3.5cm}|}{NR TDD, TR 38.901 channel model~\cite{3gpp.TR.38.901}, with UE/eNB transmission power of \SI{3}{\dBm}}                                                                                                                                                                                        \\ \cline{2-5} 
\multicolumn{1}{|l|}{}                                                & \multicolumn{3}{R{4cm}|}{UE to eNB distance (stationary UE)}                                                    & \multicolumn{1}{p{3cm}|}{$\SI{190}{\meter}, \SI{360}{\meter}, \SI{530}{\meter}, \SI{700}{\meter}$}                                                                                                                                                                     \\ \cline{2-5} 
\multicolumn{1}{|l|}{}                                                & \multicolumn{3}{R{3.5cm}|}{UE and eNB antenna height}                                              & \multicolumn{1}{l|}{$\SI{1.5}{\meter}$, $\SI{10}{\meter}$}                                                                                                                                                                                 \\ \cline{2-5} 
\multicolumn{1}{|l|}{}                                                & \multicolumn{3}{R{3.5cm}|}{ Carrier Components ([Center Frequency, Bandwidth])}                                                    & \multicolumn{1}{L{3.5cm}|}{[$\SI{28}{\giga\hertz}$, $\SI{400}{\mega\hertz}$], [$\SI{29}{\giga\hertz}$, $\SI{100}{\mega\hertz}$]} \\ \cline{2-5} 
\multicolumn{1}{|l|}{}                                                & \multicolumn{3}{R{3.5cm}|}{Network connecting the eNB to the computing node}                 & \multicolumn{1}{M{3.5cm}|}{\vspace{-0.4cm}10Gbps two-hop wired connection}                                                                                                                                                                  \\ \cline{2-5} 
\multicolumn{1}{|l|}{}                                                & \multicolumn{3}{R{3.5cm}|}{Transport protocol among splitting points}        & \multicolumn{1}{p{3.5cm}|}{TCP (NewReno, RFC 3782)}                                                                                                                                                                                           \\ \hline
\end{tabular}
\end{table}

Fig.~\ref{fig.1.2} shows the average cost difference as a function of the number of devices $|\mathcal{D}|$, in settings that are equivalent to the ones considered in Fig.~\ref{fig.1.1}. As an extension of the considerations made above, we observe that if the number of the FFNN model's layers is more than twice the number of available devices, the average cost difference increases with both $\mathcal{D}$ and $\hat{s}$. On the other hand, when the value of $|\mathcal{D}|$ approaches $|\mathcal{L}|$, the average cost difference plateaus. Once more, this was expected as, for a given value of $\hat{s}$, the more the number of devices approaches the number of FFNN model's layers, the fewer the number of ways the model's layers can be mapped onto the available devices. In turn, this simplifies the SCO problem. We remark that Fig.~\ref{fig.1} shows average performance gaps as a measure of time normalized by the term $\overline{B}/\overline{M}$. If we converted the aforementioned average performance gaps to seconds, we would need to multiply each data point by $\overline{M}/\overline{B}$, which is likely to amount to tens of \SI{}{\milli\second} for practical scenarios.

For $|\mathcal{L}| = 24$ and $|\mathcal{D}| = \{4,5,6\}$, Fig.~\ref{fig.1.2} also highlights the average time the proposed SO procedure and the SCIP solver need to solve the (linearized) version of the SCO problem, along with the average number of optimization variables and constraints forming the instances of the linearized version of the SCO problem.
When $|\mathcal{D}|$ increases from $4$ to $6$, we observe that the average number of optimization variables (constraints) increases by a factor of $2.88$ ($2.9$). Yet, the average time the SCIP solver (the proposed SO procedure) needs to solve the problem increases from $\SI{7.82}{\second}$ to $\SI{58.73}{\second}$ (from $\SI{0.14}{\second}$ to $\SI{0.35}{\second}$) -- thus, reinforcing the need of formulating a heuristic strategy.

In the remainder of this section, we will benchmark the performance of the proposed SO procedure across a selection of widely known FFNN models (summarized in Tab.~\ref{tab.I}) consisting of three Residual Neural Network (ResNet) architectures (hereafter referred to as R50, R101, and R152\footnote{ \url{https://github.com/tensorflow/tensorflow/tree/v2.15.0-rc1}.}) and versions 1-4 of You Only Look Once (YOLO) (hereafter referred to as Y1, Y2, Y3\footnote{ \url{https://github.com/samson6460/tf2_YOLO}.}, and Y4\footnote{\url{https://github.com/taipingeric/yolo-v4-tf.keras}.})~\cite{RY}. To establish a consistent cost model across the considered FFNN models, we set each model's layer CPU and memory footprint equal to the number of trainable parameters of each layer. CPU and memory normalization factors $\overline{C}$ and $\overline{M}$ have been calculated as in the case of Fig.~\ref{fig.1}.

\begin{figure}[t!]
\centering
\includegraphics[width=1\columnwidth]{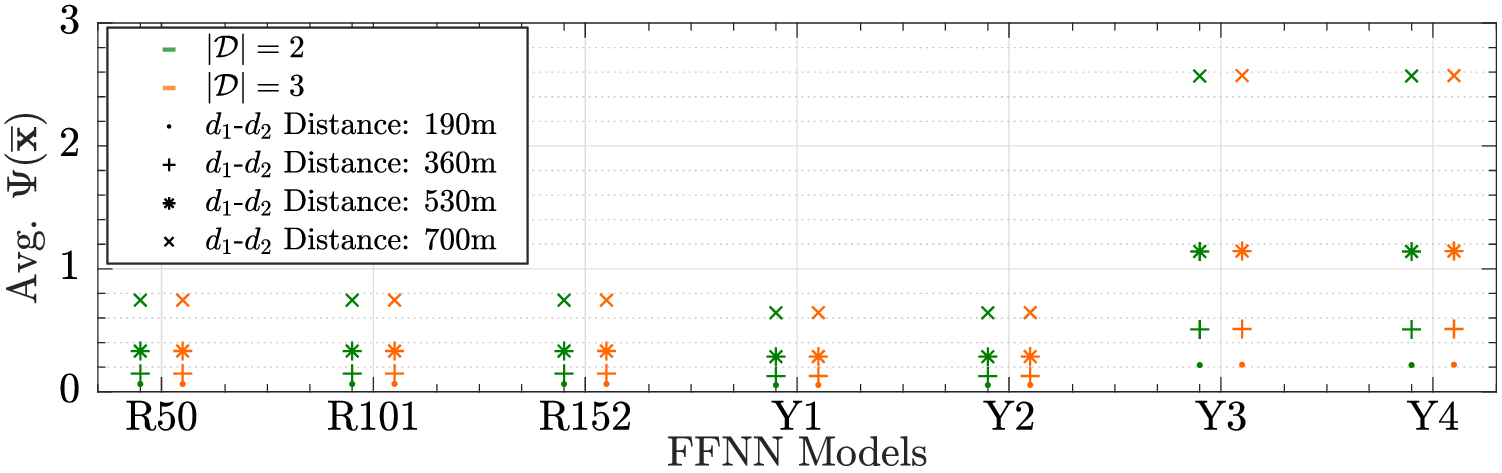}

\caption{Normalized average cost for three ResNet (R50, R101, and R152) and four YOLO architectures (Y1, Y2, Y3, and Y4), for $|\mathcal{D}| = \{2,3\}$.}
\label{fig.2.1}
\end{figure}

\begin{figure}[t!]
\centering
\hspace*{-1.5mm}\subfloat[Average Memory Footprint]{\label{fig.3.1}
    \includegraphics[width=0.495\columnwidth]{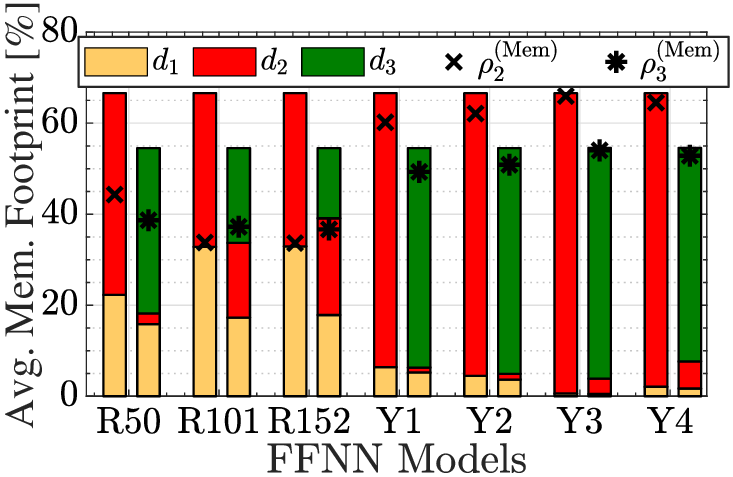}
}
\hspace*{-1.8mm}\subfloat[Average CPU Footprint]{\label{fig.3.2}
    \includegraphics[width=0.5\columnwidth]{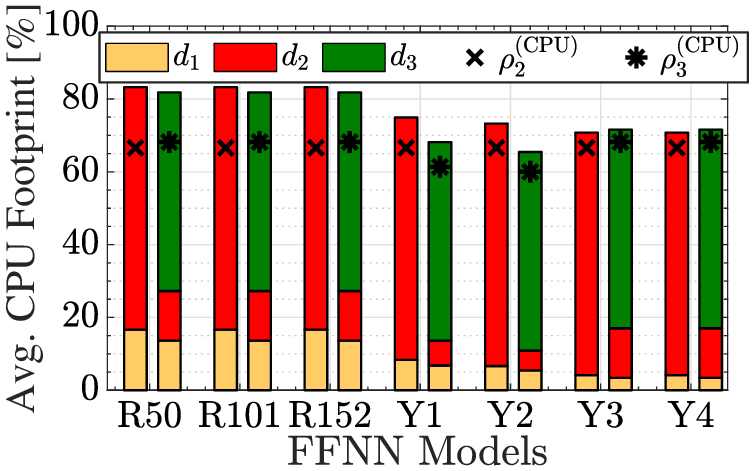}
}
\caption{Average memory and CPU footprints and UE's footprint reductions associated with the numerical results shown in Fig.~\ref{fig.2.1} for $d_1$-$d_2$ distance of \SI{700}{\meter}.}
\label{fig.3}
\end{figure}

Figs.~\ref{fig.2.1} and~\ref{fig.3} consider a system model consisting of a stationary UE ($d_1$) that is served by an eNodeB (eNB, $d_2$) by means of a 5G NR RAN. The eNB is connected to a 4G Evolved Packet Core (EPC) core network. The communication network has been simulated with ns-3's 5G-LENA~\cite{5GLENA}. The FFNN model inference task is initiated by the UE, and it may be executed in the eNB and in a dedicated computing node ($d_3$) that is ideally part of the EPC. As such, for the purpose of our performance model, the devices that can run (partially or in full) an inference in a given FFNN model can either be the UE and eNB ($|\mathcal{D}| = 2$), or the UE, eNB and a computing node part of the EPC ($|\mathcal{D}| = 3$). To account for the variability of the propagation conditions between the UE and eNB, we set $\overline{I} = 10^4$. Key simulation details are summarized in Tab.~\ref{tab.I}.

Fig.~\ref{fig.2.1} compares the (normalized) average cost of the considered ResNet and YOLO FFNN models, for four UE-eNB distances. Regardless of $|\mathcal{D}|$ and the considered FFNN model, the greater the UE-eNB, the lower the Signal-to-Interference-plus-Noise Ratio (SINR) associated with the UE's uplink communications (and measured at the eNB). In turn, the lower the SINR, the lower the modulation's order and the higher the code's overhead -- hence, the lower the value of $w_{1,2}$. For these reasons, the further the UE is from the eNB, the higher the (normalized) average cost is.
For ResNet FFNN models, the proposed SO procedure ensures a similar average cost among the models, regardless of the considered number of devices and in spite of the considerable difference in terms of the number of FFNN model's layers (see Tab.~\ref{tab.I}). This signifies that the SO procedure can efficiently establish splitting points aiming at minimizing the overall cost (subject to meeting the capacity constraints). For the YOLO FFNN models, we observe that Y1 and Y2 are relatively similar in their FFNN model structure, which leads to similar average costs. On the other hand, the number of the model's layers and the density of interconnections among layers associated with Y3 and Y4 are higher than in the case of Y1 and Y2 (see Tab.~\ref{tab.I}). This results in larger average costs for the Y3 and Y3 models. Yet, despite Y4 consisting of nearly twice as many layers as Y3, the SO procedure returns splitting solutions with similar costs.

Fig.~\ref{fig.3} compares the average memory and CPU footprints associated with results shown in Fig.~\ref{fig.2.1}, for $d_1$-$d_2$ distance of \SI{700}{\meter}. With regard to the assumption made at the beginning of this section, if $|\mathcal{D}|$ is equal to $2$ ($3$), the UE is associated with $33.3\%$ ($18.16\%$) of the overall memory and CPU capacity across the devices. In contrast, the eNB is associated with $66.6\%$ ($27.27\%$), while the dedicated computing node in the EPC is associated with $54.55\%$ (when $|\mathcal{D}| = 3$). With regards to Fig.~\ref{fig.3.1}, in the case of the ResNet models, regardless of the value of $|\mathcal{D}|$, the SO procedure tends to favor solutions associated with a larger footprint on the UE. However, as the complexity of the FFNN model increases, the SO procedure tends to return splitting solutions associated with a reduced memory footprint on the UE and/or the eNB. This is particularly evident in the case of the YOLO FFNN models -- in the case of Y4 and $|\mathcal{D}| = 3$ only $1.7\%$ and $5.9\%$ of the average memory footprint is associated with the UE and eNB, respectively. Similar considerations apply to the average CPU footprint shown in Fig.~\ref{fig.3.2}. Fig.~\ref{fig.3} also shows the average UE's memory and CPU footprint reduction ($\rho_{|\mathcal{D}|}^{(\mathrm{Mem})}$ and $\rho_{|\mathcal{D}|}^{(\mathrm{CPU})}$, respectively) defined as the average reduction of the memory/CPU footprint experienced by the UE in the case the UE had enough memory and CPU capacity to run FFNN model inferences on its own. In particular, we observe that the SO procedure ensures an average UE's memory (CPU) footprint reduction of at least $33.6\%$ or $36.6\%$ ($66.6\%$ or $60\%$) for $|\mathcal{D}|$ equal to $2$ or $3$, respectively.

{As a complementary explanation of the results shown in Fig.~\ref{fig.3}, we observe that, in the case of the considered ResNet models, the proposed SO procedure returns splitting solutions where most of the FFNN model's layers are mapped onto the UE (case $|\mathcal{D}| = 2$) or the UE and the eNB (case $|\mathcal{D}| = 3$). However, as the complexity of the FFNN model increases, the SO procedure establishes solutions where, on average, most of the model's layers are mapped onto the eNB and/or a dedicated computing node in the EPC. For instance, this is the case of Y4 where only $38.9\%$ of the layers are mapped onto the UE and the remaining onto the eNB and the dedicated computing node in the EPC, for $|\mathcal{D}| = 2$ and $3$.}

\section{Conclusions}\label{sec:conclusion}
We developed a novel optimization framework for SC applications that divides the FFNN model into sub-sets and executes them across a variable number of devices, starting from a UE and progressively considering devices further from the edge while minimizing end-to-end communication delay from wireless/wired network links.
We proposed an efficient heuristic approach to solve the formulated optimization problem. The deviation between solutions calculated with the proposed heuristic strategy and a branch-and-cut approach is marginal.
We also validated the proposed heuristic procedure on multiple state-of-the-art FFNN models with a heterogeneous number of layers, trainable parameters, and model's layer interconnections. Our numerical results show that the proposed heuristic procedure establishes valid splitting points while fulfilling device-specific computational capacity constraints.

\bibliographystyle{IEEEtran}
\bibliography{bibliography}

\end{document}